\def\icalp{0}
\documentclass[a4paper,UKenglish]{lipics-v2018}
\usepackage[utf8]{inputenc}

\usepackage{microtype}
\usepackage{graphicx,amsfonts,amsmath,amssymb,amsthm,epsfig,color,paralist,caption,subcaption,import}
\usepackage[nameinlink,capitalise]{cleveref} %
\usepackage{algorithm}
\usepackage{algpseudocode}
\usepackage[draft,footnote,nomargin,marginclue,multiuser]{fixme}
\FXRegisterAuthor{hf}{ahf}{HF}
\FXRegisterAuthor{mw}{amw}{Max}

\ifnum\icalp=1
\input{icalp}
\else
\title{A Sublinear Tester for Outerplanarity (and Other Forbidden Minors) With One-Sided Error}

\titlerunning{A Tester for Outerplanarity and Other Forbidden Minors With One-Sided Error}%

\author{Hendrik Fichtenberger}{TU Dortmund, Dortmund, Germany}{hendrik.fichtenberger@tu-dortmund.de}{https://orcid.org/0000-0003-3246-5323}{Supported by ERC grant n$^\circ$~307696.}%

\author{Reut Levi}{Weizmann Institute of Science, Rehovot, Israel}{reut.levi@weizmann.ac.il}{https://orcid.org/0000-0003-3167-1766}{Supported by ERC-CoG grant 772839.}

\author{Yadu Vasudev}{Indian Institute of Technology Madras, Chennai, India}{yadu@cse.iitm.ac.in}{https://orcid.org/0000-0001-7918-7194}{Supported by ERC grant n$^\circ$~307696.}

\author{Maximilian Wötzel}{BGSMath and UPC Barcelona, Barcelona, Spain}{maximilian.wotzel@upc.edu}{https://orcid.org/0000-0001-7591-0998}{Supported by ERC grant n$^\circ$~307696, the Spanish Ministerio de Econom\'{i}a y Competitividad projects MTM2014-54745-P, MTM2017-82166-P and an FPI grant under the Mar\'ia de Maetzu research grant MDM-2014-0445.}

\authorrunning{H. Fichtenberger, R. Levi, Y. Vasudev, M. Wötzel}%

\Copyright{Hendrik Fichtenberger, Reut Levi, Yadu Vasudev and Maximilian Wötzel}%

\subjclass{}%


\category{}%

\relatedversion{\url{http://drops.dagstuhl.de/opus/volltexte/2018/9056/}}%

\supplement{}%

\funding{}%

\acknowledgements{We would like to thank Christian Sohler for his helpful suggestions and his comments on a draft of this paper.}%

\ArticleNo{arXiv}
\nolinenumbers
\hideLIPIcs
\fi

\crefname{step}{step}{steps}
\crefname{thm}{Theorem}{Theorems}
\crefname{lem}{Lemma}{Lemmas}
\crefname{cor}{Corollary}{Corollaries}
\crefname{clm}{Claim}{Claims}
\crefname{fact}{Fact}{Facts}
\crefname{defn}{Definition}{Definitions}

\newcommand{\eps}{\epsilon}
\newcommand{\eqdef}{\stackrel{\mathrm{def}}{=}}

\newcommand{\BE}{\begin{enumerate}} \newcommand{\EE}{\end{enumerate}}
\newcommand{\BI}{\begin{itemize}} \newcommand{\EI}{\end{itemize}}
\newcommand{\calA}{{\cal A}}
\newcommand{\calC}{{\cal C}}
\newcommand{\calE}{{\cal E}}
\newcommand{\calF}{{\cal F}}
\newcommand{\calO}{{\cal O}}
\newcommand{\calB}{{\cal B}}
\newcommand{\calP}{{\cal P}}
\newcommand{\poly}{{\mathrm{poly}}}

\newtheorem{thm}{Theorem}
\newcommand{\BT}{\begin{thm}} \newcommand{\ET}{\end{thm}}
\def\FullBox{\hbox{\vrule width 8pt height 8pt depth 0pt}}
\newcommand{\ourqed}{\;\;\;\FullBox}
\newenvironment{ourproof}{\noindent\textbf{Proof:~~}}{\(\ourqed\)}
\newcommand{\BPF}{\begin{ourproof}} \newcommand {\EPF}{\end{ourproof}}
\newenvironment{proofof}[1]{\noindent\textbf{Proof of {#1}:~~}}{\(\ourqed\)}
\newcommand{\BPFOF}{\smallskip \begin{proofof}} \newcommand {\EPFOF}{\end{proofof}}
\newcommand{\BEQ}{\begin{equation}} \newcommand{\EEQ}{\end{equation}}
\newcommand{\BEQN}{\begin{eqnarray}}\newcommand{\EEQN}{\end{eqnarray}}
\newtheorem{lem}[thm]{Lemma}
\newcommand{\BL}{\begin{lem}} \newcommand{\EL}{\end{lem}}
\newtheorem{cor}[thm]{Corollary}
\newcommand{\BC}{\begin{cor}} \newcommand{\EC}{\end{cor}}
\newtheorem{clm}[thm]{Claim}
\newcommand{\BCM}{\begin{clm}} \newcommand{\ECM}{\end{clm}}
\newtheorem{fact}[thm]{Fact}
\newcommand{\BF}{\begin{fact}} \newcommand{\EF}{\end{fact}}
\newcommand{\E}{{\mathrm{E}}}
\newtheorem{defn}[thm]{Definition}
\newcommand{\BD}{\begin{defn}} \newcommand{\ED}{\end{defn}}

\newcommand{\cent}{\sigma}
\newcommand{\W}{W}  %
\newcommand{\cell}{C}  %
\newcommand{\Path}{P}  %
\newcommand{\chiW}{\chi_{\mbox{\tiny\mathit{W}}}}

\renewcommand{\l}{\ell}
\newcommand{\F}{\mathcal{F}}
\newcommand{\tr}{t}
\newcommand{\Vor}{\textsf{Vor}}
\renewcommand{\Pr}{\mathrm{Pr}}
\newcommand{\Exp}{\mathrm{Exp}}
\newcommand{\Var}{\mathrm{Var}}
\newcommand{\cov}{\mathrm{Cov}}
\newcommand{\s}[1]{\left\lvert #1 \right\rvert}
\newcommand{\e}{\epsilon}
\renewcommand{\d}{d} %
\newcommand{\dnew}[1]{{#1}}
\newcommand{\ddnew}[1]{{#1}}
\newcommand{\calT}{{\cal T}}
\newcommand{\sz}{s}  %
\newcommand{\cluster}{\textsf{cluster}}

\newcommand{\mnote}[1]{{\color{red}$\spadesuit$} \marginpar{\tiny\textbf{
            \begin{minipage}[t]{0.5in}
              \raggedright #1
         \end{minipage}}}}

\begin{document}

\maketitle

\begin{abstract}
We consider one-sided error property testing of $\calF$-minor freeness in bounded-degree graphs for any finite family of graphs $\calF$ that contains a minor of $K_{2,k}$, the $k$-circus graph, or the $(k\times 2)$-grid for any $k\in\mathbb{N}$. 
This includes, for instance, testing whether a graph is outerplanar or a cactus graph.
The query complexity of our algorithm in terms of the number of vertices in the graph, $n$,  is $\tilde{O}(n^{2/3} / \eps^5)$.
Czumaj et~al.\ showed that cycle-freeness and $C_k$-minor freeness can be tested with query complexity $\tilde{O}(\sqrt{n})$ by using random walks, and that testing $H$-minor freeness for any $H$ that contains a cycles requires $\Omega(\sqrt{n})$ queries.
In contrast to these results, we analyze the structure of the graph and show that either we can find a subgraph of sublinear size that includes the forbidden minor $H$,
or we can find a pair of disjoint subsets of vertices whose edge-cut is large, which induces an $H$-minor.
\end{abstract}

\section{Introduction}
The study of graph minors began with the work of Wagner \cite{Wag37} and Kuratowski \cite{Kuratowski1930}. 
In a seminal work on graph minors, Robertson and Seymour \cite{RS04} proved the Graph-Minor Theorem,  which states that every family of graphs that is closed under forming minors is characterized by a finite list of forbidden minors.
As a consequence of their work, they gave a classical decision algorithm with a running time of $\tilde{O}(n^3)$ to verify whether a fixed graph $H$ is a minor of $G$.

If time complexity which is polynomial, or even linear, in $n$ is considered too costly, then it is often useful to study a relaxed version of the decision problem.
In graph property testing, the goal is to test whether a given input graph has a property or is far from the property (according to some metric) while looking at a very small part of the graph (sublinear in the number of vertices).
This was first studied by Goldwasser, Goldreich and Ron~\cite{GGR98}, where the graph was represented as an adjacency matrix. %
While this model captures properties of dense graphs well, a more natural one for sparse graphs is the \emph{bounded degree} model, first studied by Goldreich and Ron \cite{GR02}, where the graph is given as adjacency lists.
In the bounded degree model, In the bounded degree model, the metric is the fraction~$\eps$ of edges of the input graph that have to be modified out of the maximum possible number of edges (the number of vertices times the bound on the maximum degree), and the \emph{query complexity} of a tester is the number of adjacency list entries that the tester looks at.
Two-sided (error) property testers are randomized algorithms that are allowed to err on all graphs, while one-sided (error) property testers are required to present a witness against the property when they reject (for more information on property testing, refer to \cite{GolInt17a,GolInt10}).

In the setting of property testing, Benjamini, Schramm and Shapira~\cite{BSS08} conjectured that for any fixed $H$, $H$-minor freeness can be tested with $\tilde{O}(\sqrt{n})$ queries by a one-sided tester on bounded degree graphs.
Czumaj et~al. \cite{CGRSSS14} provided $H$-minor freeness testers with one-sided error. They showed that $C_k$-freeness is testable with $\tilde{O}(\sqrt{n})$ queries for bounded degree graphs for $k \geq 3$ and that any one-sided tester for $H$-minor freeness requires $\Omega(\sqrt{n})$ queries when $H$ contains a cycle. When $H$ is a forest, they showed that there is a one-sided tester for $H$-minor freeness whose query complexity depends only on 
$\epsilon$. 
They consider the problem whether $H$-minor freeness can be tested with query complexity $o(n)$ for every minor $H$ “the most begging open problem”~\cite{CGRSSS14}  left open by their work.
Apart from cycles and forests, the only further progress they make towards answering this question is for $H$ which is the $4$-vertex graph consisting of a triangle and an additional edge. 
In this paper we make a significant progress on answering this question and on closing the gap between the conjecture of Benjamini et~al.~\cite{BSS08} and what is known for one-sided error testing of general forbidden minors that contain a cycle.

\subsection{Our results}
\label{sec:result}

We extend the study of testing $H$-minor freeness for a fixed graph $H$ with one-sided error. 
For a finite family of minors $\mathcal{F}$, we say that a graph is $\mathcal{F}$-minor free if it is $H$-minor free for every $H \in \mathcal{F}$.
We obtain a property tester with query complexity $\tilde{O}(n^{2/3} / \eps^5)$ for $\mathcal{F}$-minor freeness, where $\mathcal{F}$ is any family of forbidden minors that contains some graph $H$ that is a minor of either the complete bipartite graph $K_{2,k}$, the $k$-circus graph (see \cite{BodLin93}) or the $(k \times 2)$-grid. This implies, for example, that one can test with one-sided error whether a graph is outerplanar or a cactus graph.
We prove the following result (see \cref{thm:main_general} for a more general, technical version).

\BT
Given a graph $G$ with degree at most $\Delta$, and a parameter $\epsilon$, for every constant $k$, there is a one-sided (error) $\eps$-tester with query complexity and running time
$\tilde{O}(n^{2/3}/\epsilon^{5})$ for $\mathcal{F}$-minor freeness, where $\mathcal{F}$ is a family of forbidden minors such that there exists $H \in \mathcal{F}$ and $H$ is a minor of $K_{2,k}$, the $k$-circus graph or the $(k\times 2)$-grid.
\label{thm:main}
\ET

When $\mathcal{F} = \{ K_{2,3}, K_4 \}$, it follows that outerplanarity can be tested with one-sided error.
\BC
Testing outerplanarity with one-sided error has query complexity and running time $\tilde{O}(n^{2/3}/\epsilon^{5})$.
\EC

When $\mathcal{F}$ consists of the diamond graph ($C_4$ with a chord), being a cactus graph can be tested with one-sided error.
\BC
Testing the property of being a cactus graph with one-sided error has query complexity and running time $\tilde{O}(n^{2/3}/\epsilon^{5})$.
\EC

\subsection{Related Work}
As described above, the work most closely related to ours is the paper by Czumaj et~al.~\cite{CGRSSS14}. 
Other work in property testing on $H$-minor freeness studies two-sided error testers.  
Goldreich and Ron \cite{GR02} showed that $K_3$-minor freeness (i.e. cycle-freeness) can be tested with two-sided error and query complexity that is a polynomial in $1/\epsilon$ only.
Czumaj, Shapira and Sohler~\cite{CzuTes09} studied a partitioning of graphs that have low expansion (like minor free graphs), which yields two-sided tests for hereditary properties.
The problem of testing general $H$-minor freeness was studied by Benjamini, Schramm and Shapira in \cite{BSS08}, where they showed that every minor-closed property of sparse graphs is testable with two-sided error and query complexity independent of the graph's size, although they could only give an upper bound on the query complexity that was triple-exponential in $1/\epsilon$.
Hassidim et al.\ \cite{HKNO09} used partition oracles to give an easier two-sided tester for the property of $H$-minor freeness with query complexity $2^{\poly(1/\epsilon)}$. 
This was further improved by Levi and Ron \cite{LR15} to obtain a two-sided tester with query complexity that is quasi-polynomial in $1/\epsilon$. 
Yoshida and Ito~\cite{YI15} provided testers with two-sided error for outerplanarity and being a cactus graph with query complexity which is only polynomial in $1/\eps$ and the bound on the maximum degree.

\textbf{Subsequent Work.} After publication of this work, further progress was reported by Kumar et al.\ \cite{2018arXiv180508187K} who suggested an $O(n^{1/2 + o(1)})$ time algorithm to test $\mathcal{F}$-minor freeness for any family $\mathcal{F}$ of forbidden minors with one-sided error.

\subsection{Challenges and Techniques}

The results of Benjamini et~al. \cite{BSS08} imply that the optimal complexity of a two-sided $H$-minor freeness tester may depend on the size of~$H$ (and $\eps$ and $\Delta$) only.
On the other hand, it was proved by Czumaj et~al. \cite{CGRSSS14} that the hardness of the one-sided error problem depends on the structure of~$H$.
Since all embeddings of~$H$ into~$G$ may be much larger than~$H$ itself, the challenge lies in exploring the proper subgraph to find a witness.
However, finding a witness can be worthwhile because it can make the decision of the tester more comprehensible (cf. the discussion in \cite{CGRSSS14}).
In~\cite{CGRSSS14}, the problem of finding $C_k$-minors is reduced to a tester for bipartiteness~\cite{GolSub99}, which in turn finds odd cycles when two random walks that start from the same vertex collide.

Our algorithm is based on a different approach that employs a partitioning of the graph into sublinear parts.
Specifically, the main ingredients in our algorithm are a method to employ a partition of the graph that is derived from a partition into connected parts of size roughly $n^{1/3}$ by Lenzen and Levi~\cite{LL17}, and combinatorial lemmas about the existence of $H$-minors that depends on the number of cut edges between two parts.

In contrast to testing via partition oracles \cite{HKNO09,LR15}, it is not sufficient to only approximate the number of edges between the parts of our partition and, in turn, to reject if this approximated value is too high.
This would seem like a limitation to obtaining sublinear query complexity since we can no longer assume that the graph is $\Theta(\eps)$-far from being $H$-minor free after removing the edges going across the parts.
In particular, we might have to find a minor that crosses the cut of two parts.
An additional obstacle is that we cannot recover the part of a vertex in general because it can be rather large.  %
However, we can show that if the input graph is $\eps$-far from being $H$-minor free, then we can either find a large cut between two parts, which implies the existence of an $H$-minor, or we can recover a superset of a part that contains an $H$-minor.

Suppose that $G$ is $\eps$-far from being $\mathcal{F}$-minor free, where $\mathcal{F}$ is a family of forbidden minors as in \cref{thm:main}.
The algorithm uses a partition into \emph{core clusters} and \emph{remote clusters} of size $\tilde{O}(n^{1/3})$.
We draw a uniform sample of edges of constant size (which contains an edge of a minor with constant probability) and distinguish two cases.
If an edge belongs to a forbidden minor that is contained in a single cluster, then it suffices to check the cluster for this minor.
For core clusters, we can do this by using a partition oracle, but for remote clusters we use a promise on the diameter of the cluster to recover a superset of the remote cluster.

The other case is that a forbidden minor lies across clusters.
In particular, we argue that the minor must then lie across core clusters.
We show that if the cut between two clusters is greater than some threshold $f = f(H)$, this implies an $H$-minor (recall that $H$ is either $K_{2,k}$, the $k$-circus graph or the $(k \times 2)$-grid).
In fact, we show that this is true for every pair of disjoint subsets of vertices such that their respective induced subgraphs are connected. 
Given that, we analyze the edge cut of a coarser partition into \emph{super clusters} and show that if the total size of all cuts that exceed the above-mentioned threshold is small, then actually all edges between clusters can be removed. To obtain access to the partition into super clusters we make use of another coarser partition into \emph{Voronoi cells} for which we also do not have a partition oracle but, roughly speaking, can answer membership queries efficiently. 

While we do not attain the upper bound of $\tilde{O}(\sqrt{n})$ conjectured by Benjamini et al.~\cite{BSS08}, our techniques are significantly different from the ones of Czumaj et~al. \cite{CGRSSS14}, which is the only other work that gives one-sided testers for minor freeness that we are aware of.
Our work throws open two natural questions.
The first is whether the technique of partitioning can be used to obtain a one-sided tester for $H$-minor freeness that matches the conjectured $\tilde{O}(\sqrt{n})$ upper bound.
Roughly speaking, the complexity of the algorithm given in \cref{thm:main}, in terms of $n$, results from the fact that the size of the parts is $\tilde{O}(n^{1/3})$, and the fact that checking to which part a vertex belongs to takes $\tilde{O}(n^{1/3})$ as well.
The second question is whether similar techniques can be used to design one-sided testers for a larger class of minors with sublinear query complexity.
The limitation of our current approach to the aforementioned minors arises from the inner structure of the parts that we can assume, namely connectivity and bounded diameter.
Extending these guarantees, one may hope to find other minors, for example: $(k \times k)$-grid minors; or $K_{3,3}$, which implies testing planarity in sublinear time.
\section{Preliminaries}\label{sec:prel}
The graphs we consider are simple, undirected, and have a known degree bound $\Delta$.
We denote the number of vertices in the graph at hand by $n$ and we assume that each vertex $v$ has a unique id, which for simplicity we also denote by $v$.
There is a total order on the ids, i.e., given any two distinct ids $u$ and $v$, we can decide whether $u<v$ or $v<u$.
The total order on the vertices induces a total order $r$ on
the edges of the graph in the following straightforward manner:
$r(\{u,v\}) < r(\{u',v'\})$ if and only if $\min\{u,v\} < \min\{u',v'\}$ or $\min\{u,v\} = \min\{u',v'\}$
and $\max\{u,v\} < \max\{u',v'\}$.
The total order over the vertices also induces an order over those vertices visited by
a Breadth First Search (BFS) starting from any given vertex $v$, and whenever we refer to
a BFS, we mean that it is performed according to this order.
Whenever referring to one of the above orders, we may refer to the \emph{rank} of an element in the respective order.
This is simply the index of the respective element when listing all the elements according to the order starting with the smallest.

Let $G = (V,E)$ be a graph, where $V = [n]$.
We will say that a graph $G$ is $\eps$-far from a property~$P$ if at least $\eps n \Delta$ edges of $G$ have to be modified in order to convert it into a graph that satisfies the property $P$. 
In this paper, the property~$P$ that is of interest is $H$-minor freeness.
We will assume that the graph $G$ is represented by a function $f_G:[n]\times [\Delta] \to [n] \cup \{\star\}$, where $f(v,i)$ denotes the $i^{th}$ neighbor of $v$ if $v$ has at least~$i$ neighbors. Otherwise, $f_G(v,i) = \star$.
We will now define the notion of \emph{one-sided} (error) property testers.

\BD[One-sided testers]
A one-sided (error) $\eps$-tester for a property $P$ of bounded degree graphs with query complexity $q$ is a randomized algorithm ${\cal A}$ that makes $q$ queries to $f_G$ for a graph~$G$. The algorithm ${\cal A}$ accepts if $G$ has the property $P$. If $G$ is $\eps$-far from $P$, then ${\cal A}$ rejects with probability at least $2/3$.
\label{defn:one-sided}
\ED

We denote the distance between two vertices $u$ and $v$ in $G$ by $d_G(u,v)$.
For vertex $v \in V$ and an integer $r$,
let $\Gamma_r(v,G)$ denote the set of vertices at distance at most
$r$ from $v$.
When the graph $G$ is clear from the context, we shall use the shorthands
$d(u,v)$ and $\Gamma_r(v)$ for $d_G(u,v)$ and $\Gamma_r(v,G)$, respectively.
For a subset of vertices $S \subseteq V$, we denote by $G[S]$ the subgraph induced on $S$ in $G$.

\BD[Graph minors]
A graph $H$ is a minor of $G$, if $H$ can be obtained from $G$ by a sequence of vertex deletions, edge deletions and \emph{edge contractions}: For an edge $(u,v)\in G$, delete the vertices $u,v$, and create a new vertex $w$. For each neighbor $z$ of either $u$ or $v$ in the graph, add a new edge $(w,z)$.
\label{defn:minor}
\ED

\BD[$(k \times 2)$-grid]
The $(k \times 2)$-grid is the graph whose vertex set is $\{x_i\}_{i=1}^k\cup\{y_i\}_{i=1}^k$ and edge set is $\{x_i, x_{i+1}\}_{i=1}^{k-1}\cup\{y_i, y_{i+1}\}_{i=1}^{k-1}\cup\{x_i, y_i\}_{i=1}^k$.
\ED
\BD[$k$-circus graph (\cite{BodLin93})]
The $k$-circus graph is the graph whose vertex set is $\{x\}\cup\{y_i\}_{i=1}^k\cup\{z_i\}_{i=1}^k$ and edge set is $\{(y_i, z_i)\}_{i=1}^k\cup\{z_i, z_{i+1}\}_{i=1}^{k-1}\cup\{x, y_i\}_{i=1}^k$.
\ED

For a graph $G = (V,E)$ and a pair of disjoint subsets of vertices $A\subset V$ and $B\subset V$ let $E_G(A, B) \eqdef \{(u,v)\in E\,|\, u\in A \wedge v\in B\}$. When it is clear from the context, we omit the subscript.
We say that a pair of subsets of vertices $A$ and $B$ is \emph{adjacent} if $E_G(A, B) \neq \emptyset$.

\BD[Separability]
A graph $G=(V,E)$ is \emph{$(f,g)$-separable} if for every disjoint sets of vertices $A$ and $B$ such that $G[A]$ and $G[B]$ are connected and the diameter of $G[A]$ is at most $g$ it holds that $|E(A, B)| \leq f$.
\ED 

We shall use the following theorem by Erdős and Szekeres.
\BL[Erdős and Szekeres \cite{ErdCom09}]
Given a sequence of natural numbers $S = (s_i)_{i \in [n]}$ of length $n$, there exists a subsequence of length $\sqrt{n}$ that is either monotonically increasing or monotonically decreasing.
\label{lem:erdos_szekeres}
\EL

\section{Separability and Evidence for Minors}
\label{sec:comb}
\newcommand{\T}[1]{\operatorname{T}(#1)}

In this section we prove combinatorial lemmas that give sufficient conditions for the existence of a $(k \times 2)$-grid minor, a $k$-circus minor and a $K_{2,k}$ minor in a graph.
To that end, we will consider the following auxiliary graph, which is defined with respect to a partition of another graph's vertex set.

\BD
Let $G=(V,E)$ be a graph and $\mathcal{P}$ a partition of its vertex set. The graph $G[\mathcal{P}]$ is defined as the graph with vertex set $\mathcal{P}$, and $\{P,P'\} \subseteq \mathcal{P}$ is an edge if and only if there are vertices $v \in P, u \in P'$ such that $\{u,v\} \in E$. 
\ED

Notice that if $G[P]$ is connected for every $P\in \mathcal{P}$, then $G[\mathcal{P}]$ is isomorphic to the minor of $G$ obtained by contracting every edge of $G[P]$, for all $P \in \mathcal{P}$, so we will often refer to this minor also by $G[\mathcal{P}]$. 

\BL
Given a tree $T=(V,E)$ of bounded degree $\Delta$ and a subset of \emph{relevant} vertices $Q \subseteq V$, there exists a partitioning $\mathcal{P}$ of $V$ such that $T[\mathcal{P}]$ is a path minor of $T$ of length $\log_\Delta |Q|$ and for every $P \in \mathcal{P}$, it holds that $P \cap Q \neq \emptyset$.
\label{lem:minor-alg}
\EL
\BPF
We will construct the claimed partition. Initially, let $\mathcal{P} = \{\{v\} \,|\, v \in V\}$.
While there exists an edge $\{P,P'\}$ in $T[\mathcal{P}]$ %
such that $P'$ does not contain a relevant vertex and the degree of $P'$ is at most two, 
update $\mathcal{P} = \mathcal{P} \setminus P'$ and $P = P \cup P'$.
Repeat this process until no such edges remain.
Note that the resulting $T[\mathcal{P}]$ is still a tree of maximum degree $\Delta$ and that every part contains at most one relevant vertex, that is, $|P \cap Q| \leq 1$ for all $P \in \mathcal{P}$.
  
Since $T[\mathcal{P}]$ is a tree with at least $|Q|$ vertices, it has diameter $\ell \geq \log_\Delta |Q|$.
Let $L = (P_1, P_2, \ldots, P_\ell)$ be a simple path in $T[\mathcal{P}]$ between a pair of leaves, of maximum length.
For every part $P_i \in L$, let $T_i$ be the subtree rooted at $P_i$ in $T[\mathcal{P}]$ that is obtained by (virtually) removing the (at most two) edges between $P_i$ and its neighbors in $L$.
Every part $P_i \in L$ that contains no relevant vertex has at least one neighbor that is not in $L$, otherwise it is of degree at most two and would have been merged previously.
For every such part, the tree $T_i$ must contain at least one part that has non-empty intersection with $Q$, otherwise it would have been contracted.
We update the partition by setting $P_i = \bigcup_{P \in T_i} P$ for every part $i\in[\ell]$ and setting $\mathcal{P} = \{P_i\}_{i=1}^\ell$.
  
Now, we have that for every $i\in [\ell]$ , $P_i \cap Q \neq \emptyset$, and we have also not shortened the path.
Therefore, $T[\mathcal{P}]$ is the desired path minor of $T$.
\EPF

\BL
Given a rooted tree $T=(V,E)$ of bounded degree $\Delta$ with height $h$ (for some $h \in \mathbb{N}$) and a subset of \emph{relevant} vertices $Q \subseteq V$, there exists a partition, $\mathcal{P}$, of $V$ such that $T[\mathcal{P}]$ is a star minor of $T$ with $|Q| / (2h)$ leaves and for every $P \in \mathcal{P}$, it holds that $P \cap Q \neq \emptyset$.
\label{lem:minor-star}
\EL
\BPF
W.l.o.g., assume that the root is not in $Q$.
Let $S$ be the set of maximal paths in $T$ that start at the root and end at some vertex in $Q$. The length of each path in $S$ is bounded by $h$, and therefore there exist at least $|Q| / h$ such paths. Since all paths in $S$ start at the same vertex and they are maximal, their end vertices are pairwise distinct. Removing all vertices that are not contained in any path of $S$ and contracting all but the end edges of the paths in $S$, gives a star minor with at least $|Q| / h$ relevant leaves. Finally, we can contract an arbitrary leaf into the root to make the part of the root relevant.
\EPF

The common idea of the following proofs is to consider a partition of a graph into two parts such that there is a large cut, and to apply \cref{lem:minor-alg} and / or \cref{lem:minor-star} to these parts in order to construct the desired minors.

\BL
Let $G=(V, E)$ be a graph of bounded degree $\Delta$ that does not contain the $(k \times 2)$-grid as a minor.
Then, $G$ is $(\Delta^{1+\Delta^{k^2+1}}, n)$-separable.
\label{lem:cut-minor}
\EL
\BPF
We prove the contrapositive of the statement of the lemma.
Let $V_1 \dot\cup V_2$ be a partition of $V$ such that $E(V_1,V_2) > \Delta^{1+\Delta^{k^2+1}}$.
Let $T_1$ (resp. $T_2$) be a spanning tree of $G[V_1]$ (resp. $G[V_2]$).
Let $Q_1$ be the set of vertices in $V_1$ that have a neighbor in $V_2$.
Since $|E(V_1, V_2)| \geq \Delta^{1+\Delta^{k^2+1}}$, the size of the set $Q_1$ is at least $\Delta^{\Delta^{k^2+1}}$.
By \cref{lem:minor-alg}, there exists a partition $\mathcal{P}_1$ of $V_1$ such that $T_1[\mathcal{P}_1]$ is a path minor of $T_1$ of length $r \geq \log_\Delta|Q_1| \geq \Delta^{k^2+1}$. 
Let $(u_1, u_2, \cdots, u_r)$ be this path minor. 
For each vertex $u_j$ in the path minor $T_1[\mathcal{P}_1]$, it holds that $|\Gamma(u_j) \cap V_2| \geq 1$.
Now, for each vertex $u_j$ in the path minor, remove all the edges to $V_2$ except the one of lowest rank, so that $|\Gamma(u_j) \cap V_2| = 1$.

Since the degree of $G$ is at most $\Delta$, the number of vertices in $V_2$ adjacent to $Q_1$ that remain after these edge deletions is at least $\Delta^{k^2}$.
Denote this set of vertices by $Q_2$.
By \cref{lem:minor-alg}, there exists a partition $\mathcal{P}_2$ of $V_2$ such that $T_2[\mathcal{P}_2]$ is a path minor of length $t \geq \log_\Delta |Q_2| \geq k^2$.
Let $\{v_1, v_2, \cdots, v_t\}$ be this path.
Since $|\Gamma(u_j)\cap T_2[\mathcal{P}_2]| =1$ for all $u_j \in T_1[\mathcal{P}_1]$, we have $t \leq r$. Furthermore, each vertex $v_j \in T_2[\mathcal{P}_2]$ has at least one neighbor in $T_1[\mathcal{P}_1]$.
Therefore, by Hall's theorem, there is matching of size at least $k^2$ between $T_1[\mathcal{P}_1]$ and $T_2[\mathcal{P}_2]$. 

By \cref{lem:erdos_szekeres}, there is a set of $k$ vertices, say $u_{i_1}, u_{i_2}, \cdots, u_{i_k}$ and $v_{j_1}, v_{j_2}, \cdots, v_{j_k}$, such that $i_1 \leq i_2 \leq \cdots \leq i_k$ and $j_1 \leq j_2 \leq \cdots \leq j_k$, and $(u_{i_l},v_{j_l})$ is an edge in the matching. Contract the edges in the path to the vertices $u_{i_1}, u_{i_2}, \cdots, u_{i_k}$ and $v_{j_1}, v_{j_2}, \cdots, v_{j_k}$ to obtain the $(k\times 2)$-grid minor.
\EPF

\BL
Let $G=(V,E)$ be a graph of bounded degree $\Delta$ that does not contain the $k$-circus as a minor.
Then, $G$ is $(2h \Delta^{2+k^2},h)$-separable for every $h \in \mathbb{N}$.
\label{lem:cut-circus}
\EL
\BPF
We prove the contrapositive of the statement of the lemma.
This proof is very similar to the proof of \cref{lem:cut-minor}. The only difference is that~(i) we consider a partition $V_1 \dot\cup V_2$ of $V$ such that $E(V_1,V_2) > 2h \Delta^{2+k^2}$ and the diameter of $G[V_1]$ is at most $h$ and~(ii) we apply \cref{lem:minor-star} instead of \cref{lem:minor-alg} to $G[V_1]$.
In particular, let $T_1$ (resp. $T_2$) be a spanning tree of $G[V_1]$ (resp. $G[V_2]$).
Let $Q_1$ be the set of vertices in $V_1$ that have a neighbor in $V_2$.
Since $|E(V_1, V_2)| \geq 2h \Delta^{2+k^2}$, the size of the set $Q_1$ is at least $2h \Delta^{1+k^2}$.
By \cref{lem:minor-star}, there exists a partition $\mathcal{P}_1$ of $V_1$ such that $T_1[\mathcal{P}_1]$ is a star minor of $T_1$ with $r \geq \Delta^{1+k^2}$ leaves.
Let $\{u_1, u_2, \cdots, u_r\}$ be the leaves of this star minor. 
For each leaf $u_j$ of the star minor $T_1[\mathcal{P}_1]$, it holds that $|\Gamma(u_j) \cap V_2| \geq 1$.
Now, for each leaf $u_j$ in the star minor, remove all the edges to $V_2$ except the one of lowest rank, so that $|\Gamma(u_j) \cap V_2| = 1$.
The remaining proof is analogous to the proof of \cref{lem:cut-minor}.
\EPF

\ifnum\icalp=1
The proof of the following lemma appears in the arXiv version~\cite{FicSub17}.
\fi
\BL
Let $G=(V,E)$ be a graph of bounded degree $\Delta$ that does not contain the complete bipartite graph $K_{2,k}$ as a minor.
Then, $G$ is $(2 \Delta kh,h)$-separable for every $h \in \mathbb{N}$.
\label{lem:cut-K2k}
\EL

\def\comlem{
Observe that $k$-circus contains $K_{2,k}$ as a minor, and therefore it follows from \cref{lem:cut-circus} that $G$ is $(2h \Delta^{2+k^2},h)$-separable. However, a better bound can be achieved directly as follows: instead of applying \cref{lem:minor-alg} to $G[V_2]$ in the proof of \cref{lem:cut-circus} (that is, carrying out the shared part of the proofs of \cref{lem:cut-minor,lem:cut-circus}), one can simply contract $V_2$ to a single vertex to obtain $K_{2,k}$.
}

\ifnum\icalp=0
\BPF
\comlem
\EPF
\fi

\section{Underlying Partitions}\label{sec:part}
In this section we describe a method to partition the graph into small connected parts with certain properties that enable us to apply \cref{lem:cut-minor,lem:cut-circus,lem:cut-K2k}.
The partition technique is very similar to the one that appears in~\cite{LL17}, which is used for the local construction of sparse spanning subgraphs.
We make minor adaptations to suit our needs. 
\ifnum\icalp=1
Omitted proofs appear in the arXiv version~\cite{FicSub17}.
\fi
In \cref{sec:alg}, we will show how to utilize these partitions for testing the forbidden minors.   

Three different partitions are described next, one of which is a refinement of the other two.
As described in more detail in the next sections, the properties of these partitions are as follows. The refined partition into \emph{core clusters} can be locally recovered.
The edge cut of this partition is not necessarily small, even if the input graph excludes the forbidden minor. 
The second partition into \emph{Voronoi cells} will be useful for checking the edge cut of the third partition into \emph{super clusters}, which in turn is guaranteed to have a small edge cut in case the graph excludes the forbidden minor. 
\ifnum\icalp=1
See the arXiv version~\cite{FicSub17} for an illustration of the following definitions.
\else
See \cref{fig:partitioning} for an illustration of the following definitions.
\fi

\textbf{\textsf{Parameters.}} 
The input parameters are $\alpha$ and $\gamma$. 
We sample $\ell$ uniformly at random from $[b\log n/ \log(1+ \gamma),b\log n/ \log(1+ \gamma) + \Delta/\gamma]$, and let $t \eqdef cn^{1/3} \ln n \cdot \ell\Delta/\alpha$ where $c$ and $b$ are sufficiently large constants. 
The parameter $\ell$ affects the diameter of the parts of the partition. It is picked randomly so as to ensure that only a small fraction of the edges are in the edge cut of the partition. 

\textbf{\textsf{Centers.}} Pick a set $S\subset |V|$ of $\Theta(\alpha n^{2/3}/\ln n)$ vertices at random. 
We shall refer to the vertices in $S$ as \emph{centers}.  
For each vertex $v \in V$, its \emph{center}, denoted by $c(v)$, is the center which is closest to $v$ among all centers (break ties between centers according to the rank). 

\textbf{\textsf{Remote Vertices.}} 
Define $R \eqdef \{v \,|\, \Gamma_\ell(v) \cap S = \emptyset\}$ where $S$ is the set of centers.
We call the vertices in $R$ \emph{remote} and abbreviate $\bar{R}\eqdef V\setminus R$.

\textbf{\textsf{Voronoi cells.}} The \emph{Voronoi cell} of a vertex $v \in \bar{R}$ is $\Vor(v)  \eqdef \{u \in \bar{R} \,|\, c(u) = c(v)\}$.

\medskip
We deal with the partitioning of remote vertices later.
Given a vertex $v \in \bar{R}$, one can determine its center by exploring its $\ell$-hop neighborhood.
However, it is much more costly to find all vertices that belong to its Voronoi cell, which may have size $\Omega(n)$.
We now describe how to further refine the partition given by the Voronoi cells so that the number of vertices in each cluster is $\tilde{O}(n^{1/3} \Delta / \alpha)$.

\textbf{\textsf{Core clusters.}}
For each Voronoi cell, consider the BFS tree spanning it, as described in \cref{sec:prel}, which is rooted at the respective center.
For every $v\in V$, let $p(v)$ denote the \emph{parent} of $v$ in this BFS tree. If $v$ is a center then $p(v) = v$.
For every $v\in V\setminus S$, let $T(v)$ denote the subtree of $v$ in the above-mentioned BFS tree when we remove the edge $\{v, p(v)\}$.
Now consider a Voronoi cell.
We define the \emph{core cluster} of a vertex $v$ as follows:
\begin{enumerate}
\item If $|\Vor(v)| \leq t$ then the core cluster of $v$ is $\Vor(v)$.
\item If $|T(v)| \geq t$, where $|T(v)|$ denotes the number of vertices in $T(v)$, then the core cluster of $v$ is the singleton $\{v\}$.
\item Otherwise, $v$ has a unique ancestor $u$ for which $|T(u)| < t$ and $|T(p(u))| \geq t$. 
The core cluster of $v$ is the set of vertices in $T(u)$. 
\end{enumerate}
For a vertex $v \in \bar{R}$, let $\cluster(v)$ denote the cluster of $v$.
For a cluster $C$, let $c(C)$ denote the center of the vertices in $C$ (all the vertices in the cluster have the same center).
Let $\Vor(C)$ denote the Voronoi cell of the vertices in $C$. 

\def\figcen{
\begin{figure}
	\centering
	\begin{subfigure}[b]{88mm}
		\centering
		\includegraphics[width=88mm]{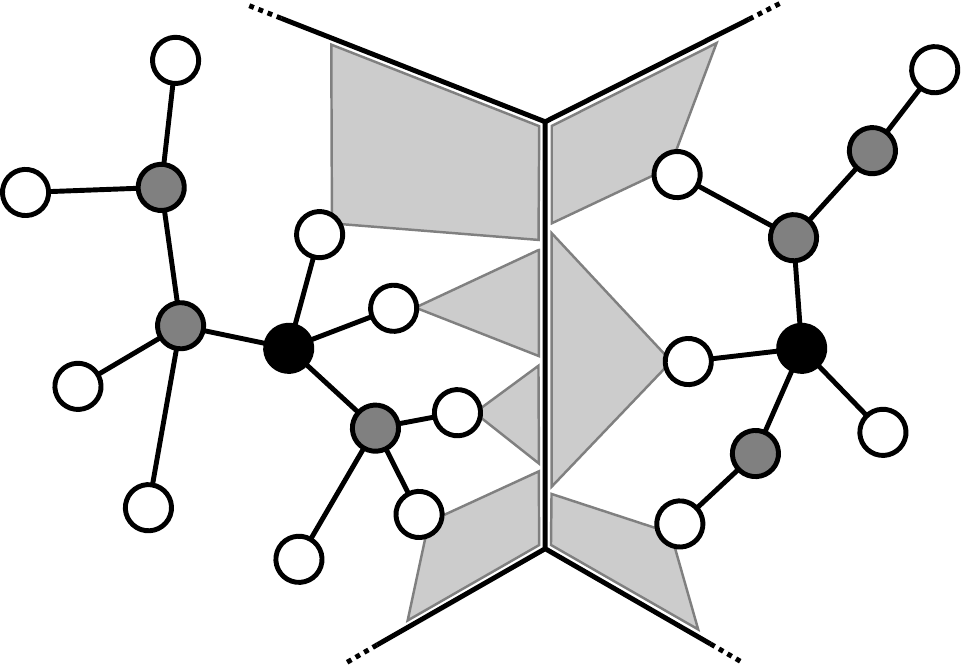}
		\caption{\emph{Voronoi cells and core clusters.} Centers (black) constitute the voronoi cells. Vertices with more than $t$ children in their BFS~subtrees are singletons (gray). Every other vertex and the vertices in its subtree form a core cluster (white, subtrees simplified / omitted).}
		\label{fig:voronoi_cells}
	\end{subfigure}
	\hspace{10pt}
	\begin{subfigure}[b]{45mm}
		\centering
		\includegraphics[width=45mm]{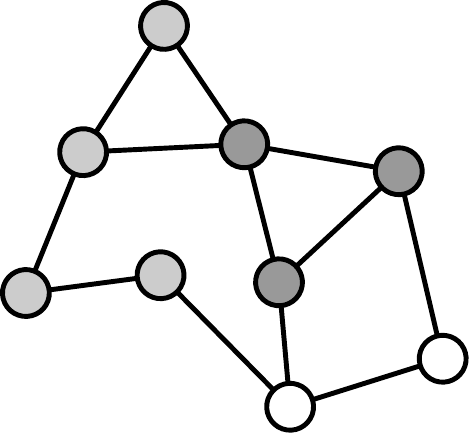}
		\caption{\emph{Remote clusters.} Assume that $r_y \leq 5$ for $y \notin \{u,v,w\}$ and that the ranks are $u < v < w$. Then, $m_v(x) = m_w(x) = 7, m_u(x) = 6$ and $x$ belongs to the remote cluster of $v$ (white) because it has lower rank than $w$. Furthermore, $C(x) = \{u,v,w\}$.}
		\label{fig:remote_clusters}
	\end{subfigure}
	\caption{Partitioning a graph into Voronoi cells, core clusters and remote clusters.}
	\label{fig:partitioning}
\end{figure}
}

\ifnum\icalp=0
\figcen
\fi

\medskip
This describes a partition of $V$ into $R$ and  $\bar{R}$, a refinement of $\bar{R}$ into Voronoi cells, and a refinement of this partition into core clusters\ifnum\icalp=0 (see \cref{fig:voronoi_cells})\fi.
It was shown in~\cite{LL17} that the number of core clusters is not much higher than the number of Voronoi cells.
\BL[Lemma 1 in~\cite{LL17}]
\label{lem:bound_s}
The number of core clusters, denoted by $s$, is at most $|S| + n \ell (\Delta + 1)/t$.
\EL

Note that core clusters are, like Voronoi cells, connected.

\BL
For every vertex $v \in \bar{R}$, $\cluster(v)$ is connected.
\label{lem:connected1}
\EL

\def\connlem{
This follows from the construction since every core cluster is either a Voronoi cell, a singleton vertex, or the subtree of the BFS tree of a Voronoi cell.
}	

\ifnum\icalp=0
\BPF
\connlem
\EPF
\fi

Even more, Voronoi cells are still connected if one removes a cluster that is not a singleton.

\BL
Let $v \in \bar{R}$ be a vertex such that $\cluster(v)$ is not a singleton. Then, $G[\Vor(v) \setminus \cluster(v)]$ is connected.
\label{lem:connected2}
\EL

\def\connseclem{
Observe that if $\cluster(v)$ is not a singleton, then it is the subtree of the BFS tree of $G[\Vor(v)]$ rooted at its center. Therefore, removing $\cluster(v)$ does not disconnect $G[\Vor(v)]$.
}

\ifnum\icalp=0
\BPF
\connseclem
\EPF
\fi

In contrast to Voronoi cells, core clusters are guaranteed to be sufficiently small by construction, which allows us to fully explore them in an efficient manner. An explicit procedure for this is given in \cref{sec:local_imp}. However, it might still be possible for the overall edge cut to be large, even if there are only few edges in individual cuts between two core clusters.
To this end, we group core clusters and consider the cut between pairs of a core cluster and such a resulting \emph{super cluster} instead.

\BD
For a core cluster $A$, define its \emph{adjacent vertices} $\partial A\eqdef \{v\,|\,u\in A \wedge v\in \bar{R}\setminus A \wedge \{u,v\}\in E \}$, i.e., the set of vertices  that are adjacent to a vertex in $A$, excluding $A$. 
\ED
\BD
Define the \emph{adjacent centers} of a set of vertices $A \subseteq \bar{R}$ to be $\{c(v)\,|\,v\in A\}$.
\ED

\textbf{\textsf{Marked Clusters.}}
Each center is \emph{marked} independently with probability $p \eqdef 1/n^{1/3}$. 
If a center is marked, then we say that its Voronoi cell is marked and all the clusters in this cell are marked as well.

\textbf{\textsf{Super Clusters.}}
Let $A$ be a cluster which is not marked but is adjacent to at least one marked cluster. 
Let $\{u, v\}$ be the edge with minimum rank such that $u\in A$ and $v \in B$, where $B$ is a marked cluster.
We say that the cluster $A$ \emph{joins} the cluster $B$. The \emph{super cluster} of $B$ consists of $B$ and all the clusters which join $B$. 

\medskip
After considering pairs of clusters and super clusters (and bounding the number of edges between them), only few pairs of core clusters $(A,B)$ are left such that neither $A$ nor $B$ are member of a super cluster with high probability.

\BL
With probability at least $1-o(1)$, it holds that $|c(\partial A)| \leq n^{1/3}\log n$ for every core cluster $A$ that is not adjacent to a marked cluster.
\label{lem:marked}
\EL

\def\markedlem{
The probability that cluster $A$ is not adjacent to a marked cell is $(1-p)^{|c(\partial A)|}\leq e^{-p|c(\partial A)|}$. Therefore, if $|c(\partial A)| > p^{-1}\ln n$, then the probability that $A$ is not adjacent to a marked cluster is at most $1/n$. Since there are at most $\tilde{O}(n^{2/3})$ many core clusters, the lemma follows from a union bound.
}	

\ifnum\icalp=0
\BPF
\markedlem
\EPF
\fi

This settles the three partitions of vertices from $\bar{R}$ into Voronoi cells, core clusters and super clusters.
We now describe a way to partition the remote vertices into remote clusters \ifnum\icalp=0 (see \cref{fig:remote_clusters})\fi such that (with high probability) the total number of edges that go out from each remote cluster is at most $O(\gamma n \Delta)$ even if the graph is far from being $H$-minor free.
Basically, this implies that one can test a remote cluster isolated from the remaining graph because all outgoing edges can be removed such that $G$ which was $(\alpha + \gamma)$-far from being $H$-minor free is still $\alpha$-far from the property.
The partitioning uses ideas of Elkin and Neiman~\cite{EN17}.

\textbf{\textsf{Remote clusters.}}
We will first describe the algorithm of Elkin and Neiman~\cite{EN17}.
Given an integer $h$ and a parameter $0 < \delta \leq 1$, each vertex $v$ draws $r_v$ according to the exponential distribution with parameter $\beta = \ln (n/\delta)/h$. By Claim 2.3 in~\cite{EN17}, with probability at least $1-\delta$, it holds that $r_v < h$ for all $v \in V$.
Each vertex $v$ receives $r_u$ from every vertex $u$ within distance at most $h$, and stores the values $m_u(v) = r_u - \d(u ,v)$. 
We use this technique to obtain a partition of $R$ as follows.
Every vertex $v\in R$ is \emph{assigned} to the vertex $u \in R$ such that $m_u(v) = \max_{w\in R}\{m_w(v)\}$, if there is more than one such vertex, pick the one with minimum rank\ifnum\icalp=0 (see \cref{fig:remote_clusters})\fi.
We say that $u$ is the \emph{leader} of $v$ denoted by $L(v)$ and that $\{w \in R \,| L(w) = L(v)\}$ is the \emph{remote cluster} of $v$.
We note that we run this algorithm on $G[R]$ (namely, we calculate $m_u(v) = r_u - \d_{G[R]}(u ,v)$), therefore a vertex $u$ can not be assigned to a vertex on a different connected component in $G[R]$. 

Like core clusters, remote clusters are also connected.

\BL \label{lem:connected}
For every $v \in R$, the subgraph induced on the remote cluster of $v$ is connected. 
\EL 

\def\connthlem{
Consider any shortest path between $v$ and $v' \eqdef L(v)$ in $G[R]$.
Let $w$ be the neighbor of $v$ on this shortest path. We claim that $L(w)= L(v)$. From this the lemma follows by induction on the distance to $L(v)$.
Assume to the contrary that $w' \neq v'$ where $w' \eqdef L(w)$. 
Then either $m_{w'}(w) > m_{v'}(w)$ or, $m_{w'}(w) = m_{v'}(w)$ and  the rank of $w'$ is higher than the rank of $v'$. 
Since $m_{v'}(v) = m_{v'}(w) -1$ and $m_{w'}(v) \geq m_{w'}(w) -1$ we obtain that in both cases $w'$ is the leader of $v$, contrary to our assumption.
}

\ifnum\icalp=0
\BPF
\connthlem
\EPF
\fi

Similarly\footnote{This set is defined slightly differently in~\cite{EN17}, but \cref{Elk.lem} applies to this definition as well.} as in~\cite{EN17}, we define $C(v) = \{u\,|\, m_u(v) \geq \max_{w\in V}\{m_w(v) -1\}\}$, for every $v\in R$. 
We will use an observation about the size of this set to argue that the number of cut-edges between remote clusters is small.
\BL[Proof of Lemma 2.2 in~\cite{EN17}]\label{Elk.lem}
For every $v\in R$, $\Exp[C(v)] \leq (n/\delta)^{1/h}$.
\EL
For $\delta = 1/n^{b-1}$ and $h = \ell$, we obtain that $\Exp[C(v)] \leq (1+\gamma)$.
Define the \emph{edge-cut of $R$} to be $K = \{ \{u, v\} \in E \,| v \in R\ \wedge u \in R \wedge (L(u) \neq L(v))\}$.

\BL\label{H.lem}
With probability at least $99/100$, $|K| \leq 100 \gamma \Delta |R| \leq 100 \gamma \Delta n$.  
\EL

\def\hlem{
Define $B = \{v\in R \,| C(v) > 1\}$. 
Observe that $e \cap B \neq \emptyset$ for any edge $e\in K$.
To see this, consider an edge $\{u, v\} \in K$. Let $u'$ and $v'$ denote the leaders of $u$ and $v$, respectively, 
and assume w.l.o.g. that $u'$ has higher rank than $v'$.
By the triangle inequality $\d(v', u) \leq \d(v', v) + 1$, thus we have that $m_{v'}(u) \geq m_{v'}(v) - 1$.
Since $u'$ has higher rank than $v'$ it holds that $m_{u'}(u) \leq m_{v'}(v)$, and so it follows that $m_{v'}(u) \geq m_{u'}(u) - 1$.
Thus $\{v', u'\} \subseteq C(u)$, which implies that $u\in B$, as required.  

Hence $|K| \leq \Delta |B|$.
By \cref{Elk.lem} and the linearity of expectation, we obtain that with probability at least $99/100$, $|B| \leq 100\gamma |R|$ and so we get the desired bound on $|K|$.
}

\ifnum\icalp=0
\BPF
\hlem
\EPF
\fi

It remains to bound the number of edges between remote clusters and core clusters.

\BL[Lemma 6 in~\cite{LL17}]\label{rr.lem}
$\Exp[|E(R,\bar{R})|]\leq \gamma n$.
\EL

\section{The Algorithm}\label{sec:alg}

We  prove the following result.
\Cref{thm:main} follows by plugging in \cref{lem:cut-minor,lem:cut-circus,lem:cut-K2k}.

\BT
\label{thm:main_general}
Let $\calF$ be a finite family of graphs such that there exists $H \in \calF$ and $f=f(n,\calF,\Delta),\, g=g(n,\calF,\Delta), g\geq\ell$ (where $\ell$ is defined in \cref{sec:part}) such that every $n$-vertex graph that is $H$-minor free is $(f,g)$-separable. For every $\epsilon > 0$, there is a one-sided $\eps$-tester that given query access to an $n$-vertex graph, $G$, with maximum degree $\Delta$, tests whether $G$ is $\calF$-minor free (i.e., $G$ is $R$-minor free, for every $R \in \calF$).  
The query complexity of the tester is $\tilde{O}(n^{2/3} f^3 / \epsilon^{5})$. If $\calF$ includes a planar graph, then the running time is  $\tilde{O}(n^{2/3} f^3 / \epsilon^{5})$ as well.
\ET

We first analyze a global version of the tester (see \cref{alg.main}) and show how to turn it into a local algorithm in \cref{sec:local_imp}.
Our tester draws $\Theta(f/\eps)$ edges at random from the input graph~$G$.
It follows that at least one of these edges is part of a forbidden minor with constant probability if~$G$ is $\eps$-far from being $\mathcal{F}$-minor free.
For the sake of this exposition, think of the graph being partitioned into core clusters and remote clusters.
To reveal a forbidden minor from $\mathcal{F}$, the algorithm employs this partitioning.

We conduct the following case analysis.
Either, an edge that is one out of many that connect a cluster $A$ and an adjacent (disjoint) super cluster $B$ will be sampled.
Since non-remote clusters have diameter at most $\ell \leq g$, by the $(f, g)$-separability of $\mathcal{F}$-minor free graphs, a large cut between $A$ and $B$ implies the existence of $H$ as a minor (see \cref{step.samevor,step.notsamevor}).
Otherwise, one can show that the total number of edges between clusters is at most $\eps n \Delta / 2$.
This implies that the edges between clusters can be removed such that the graph is still $\eps / 2$-far from being $\mathcal{F}$-minor free.
Then, it suffices to look for a minor-instance of a graph from $\mathcal{F}$ inside the clusters of each edge (see \cref{step.algmain}).

Therefore, it suffices to bound the number of edges between clusters under the promise that the edge-cut between every cluster and an adjacent (disjoint) super cluster is small.
Since a naive bound over all pairs of clusters is quite costly, we classify edges and analyze each class independently.
First, we observe that the total number of edges between remote clusters and clusters is $O(\eps n \Delta)$ with constant probability. It remains to bound the number of edges between core clusters.
To this end, we analyze the total number of edges between core clusters within the same Voronoi cell, the total number of edges between two unmarked core clusters and the total number of edges between core clusters and super clusters separately.
This covers all relevant edges between clusters at least once and gives an upper bound on their total number.

\Cref{thm:main_general} follows from the the efficient implementation (\cref{sec:local_imp}) and the correctness of the tester (\cref{sec:correctness}). 

\def\noteon{
\paragraph*{Note on the running time} The running time of our $\calF$-minor testing algorithm is also $\tilde{O}(n^{2/3}f^3/\eps^5)$, provided that $\calF$ contains a planar graph $H'$.
In Steps \ref{step.samevor}, \ref{step.ecv} and \ref{step.cu}, Algorithm \ref{alg.main} rejects based on the number of edges crossing a cut. 
The time complexity for these steps is the same as the query complexity to compute the cut size. 
In~\cref{step.algmain}, \cref{alg.main} tests if any graph in $\calF$ exists as a minor in the cluster $C$ of size at most $t=\tilde{O}(n^{1/3}f/\eps^2)$.
To check this we proceed as follows.
Let $m = 2|V(H')|+4|E(H')|$.
Now check whether $G[C]$ has treewidth at most $20^{2m^5}$. 
This can be done in time $O(t)$ (see \cite{Bod96}). 
By a theorem of Robertson, Seymour and Thomas \cite{RobertsonST94}, if the treewidth is more than $20^{2m^5}$, then $G[C]$ contains $H'$ as a minor and the algorithm can reject right away. 
Otherwise, we know that $G[C]$ has treewidth at most $20^{2m^5}$.
Now, there is an $O(t)$-time algorithm by Courcelle to test if $G[C]$ contains $H$ as a minor \cite{Courcelle90}.
}

\ifnum\icalp=1
The proof of the running time appears in the arXiv version~\cite{FicSub17}.
\else
\noteon
\fi

\begin{algorithm}
\caption{Test $\mathcal{F}$-minor freeness}\label{alg.main}
\BE
\item Partition $V$ according to the partition described in Section~\ref{sec:part} with parameters $\gamma= \Theta(\eps)$ and $\alpha=\Theta(\eps/f)$.
\item Sample $\Theta\left(f / \eps \right)$ random edges from $G$. For each sampled edge $\{u, v\}$:
\BE 
\item Find the cluster for each endpoint, denoted by $C_v$ and $C_u$, respectively. 
\item If both $u$ and $v$ belong to the same cluster $C$, check that $G[C]$ is $\mathcal{F}$-minor free, if it is not, then return \textrm{REJECT}.\label[step]{step.algmain}
\item If either for $w = u$ or for $w = v$  it holds that: $w\in \bar{R}$, $\cluster(w)$ is not a singleton, and $|E(\Vor(w) \setminus \cluster(w), \cluster(w))| > f$, then return \textrm{REJECT}.\label[step]{step.samevor}
\item If both $u$ and $v$ are in $\bar{R}$ and and both $C_u$ and $C_v$ are not singletons then: \label[step]{step.notsamevor}
\BE
\item If $|E(C_v, C_u)| > f$ then return \textrm{REJECT}.\label[step]{step.ecv}
\item If $C_v$ (and symmetrically for $C_u$) joins a cluster $C \neq C_u$, then let  $A \eqdef \bigcup_{v \in \partial C\setminus C_u} \Vor(v)$. If $|E((A \cup C) \setminus C_u, C_u)|> f$ then return \textrm{REJECT}. \label[step]{step.cu}
\EE
\EE
\item Return \textrm{ACCEPT}.
\EE 
\end{algorithm}

\subsection{Efficient Implementation}
\label{sec:local_imp}
In this subsection we describe how \Cref{alg.main} can be implemented in query \ifnum\icalp=0 and time \fi complexity $\tilde{O}(n^{2/3}f^3/\eps^5) \cdot \poly(\Delta)$.  
For a vertex $v \in V$, define $i_v \eqdef \min_{i}  \{ \Gamma_{i}(v) \geq y\}$ where $y \eqdef \Theta(n^{1/3}\log^2 n/ \alpha)$.  
Let $E$ denote the event that $\Gamma_{i_v}(v) \cap S \neq \emptyset$ for all $v \in V$. 
Since w.h.p. $E$ occurs, henceforth we condition on this event. 
The following subroutines are sufficient in order to implement~\Cref{alg.main}:
\begin{enumerate}
\item Given a vertex $v \in \bar{R}$, finding $\cluster(v)$. The query complexity of finding $c(v)$ is bounded by $y \cdot \Delta$.
That is, $c(v)$ is found after performing a BFS from $v$ for at most $i_v$ levels.
Furthermore, the path connecting $v$ and $c(v)$ in $T(c(v))$ is also found (this is the shortest path between $v$ and $c(v)$ of smallest lexicographical order).
Therefore it is possible to explore $T(c(v))$ with query complexity $O(y \cdot \Delta)$ per step.
In order to determine $\cluster(v)$, it is sufficient to explore $T(v)$ and $T(a)$ up to $t$ vertices, for any ancestor, $a$, of $v$.
Therefore, the total query complexity is at most $O(\ell t y \Delta) = \tilde{O}(n^{2/3}f^2/\eps^{4}) \cdot \poly(\Delta)$ per iteration.   
\item Given a subset of vertices $A$, finding $c(\partial A)$ can be done in query complexity $y \cdot \Delta$ times the total number of edges which are incident to vertices in $A$. If $A$ is a cluster then the latter is bounded by $t \Delta$. Therefore we obtain a bound of $\tilde{O}(n^{2/3}f^2/\eps^3) \cdot \poly(\Delta)$ queries per iteration of \cref{step.samevor} and \cref{step.cu}.
\item Finally, instead of finding the remote-cluster of a vertex $v \in R$ it suffices to find a connected induced subgraph that contains the remote-cluster of $v$. This is achievable by first finding the corresponding leader and then exploring the $\ell$-hop neighborhood of the leader. To find the leader, it is sufficient to explore the $\ell$-hop neighborhood of $v$ and then for every vertex in it, to determine whether it is in $R$ or not (determining whether a vertex is in $R$ takes $O(y \cdot \Delta)$ time).  
With this at hand, it is possible to simulate the result of the leader-decomposition algorithm for $v$. Observe that both $v$ and the leader of $v$ are in $R$, therefore (under the assumption that $E$ occurred) it is possible to explore their $\ell$-hop neigberhood in $O(y \cdot \Delta)$ time. The query complexity for each iteration of this step is bounded by $\tilde{O}(n^{2/3}f^2/\eps^2)\cdot \poly(\Delta)$.
\end{enumerate}

\subsection{Correctness}
\label{sec:correctness}

\BL
\Cref{alg.main} accepts every graph $G$ that is $\mathcal{F}$-minor free.
\label{lem:completeness}
\EL
\BPF
The completeness of the test is based on the separability of $H$-minor free graphs. We show that if the algorithm rejects, then $G$ contains a graph from $\mathcal{F}$ as minor.

\Cref{step.algmain} rejects only if $G$ contains a graph from $\mathcal{F}$ as a minor.
To apply $(f, g)$-separability to \cref{step.samevor}, it suffices to note that for any $w\in \bar{R}$ such that $\cluster(w)$ is not a singleton, $G[\Vor[w] \setminus \cluster(w)]$ is connected by \cref{lem:connected2}.

To apply the separability to \cref{step.ecv}, it suffices to note that $C_v$ and $C_u$ are disjoint and that $G[C_v]$ and $G[C_u]$ are both connected by \cref{lem:connected1}.
To apply the separability to \cref{step.cu}, we need to show that $G[(A\cup C)\setminus C_u]$ is connected, since it is clearly disjoint from $C_u$, the correctness then follows.
There are two cases. If $(A\cup C) \cap C_u = \emptyset$, then $G[(A\cup C)\setminus C_u] = G[A\cup C]$ is connected.
Otherwise, since $C_u$ is not a singleton and since $\partial C$ contains a vertex $v$ which is in $\Vor(C_u)\setminus C_u$, the claim follows from~\cref{lem:connected2}.
\EPF

\BL
\Cref{alg.main} rejects every graph $G$ that is $\eps$-far from being $H$-minor free with probability $2/3$.
\label{lem:soundness}
\EL
\BPF
Assume that $G$ is $\eps$-far from being $H$-minor free.
Let $\calP$ denote the partition obtained by the algorithm (namely, the partition of the entire graph as described in \cref{sec:part} with parameters $\gamma= \Theta(\eps)$ and $\alpha=\Theta(\eps/f)$).
We say that an edge $e = \{u ,v\}$ violates the separability property with respect to $\calP$ if either:
\begin{enumerate}
\item There exist a core cluster $A \in \bar{R}$ and a cluster or a super cluster, $B \in \bar{R}$ such that $e \in E(A, B)$ and $|E(A, B)| > f$,
\item or, if either for $w = u$ or for $w = v$  it holds that: $w\in \bar{R}$, $\cluster(w)$ is not a singleton, and $|E(\Vor(w) \setminus \cluster(w), \cluster(w))| > f$.
\end{enumerate} 
Let $\calE$ denote the set of edges which violate the $f$-separability property with respect to $\calP$.
If $|\calE| > \alpha n \Delta$, then with probability at least $99/100$, the algorithm finds a violation in one of the steps:~\cref{step.samevor},~\cref{step.ecv} or~\cref{step.cu}.
Note that we do not need to check any edges between a core cluster and a remote cluster nor any edges between two remote clusters.
By Markov's inequality and~\cref{rr.lem}, with probability at least $99/100$, $|E(R, \bar{R})| \leq 100 \gamma n$.
By~\cref{H.lem}, with probability at least $99/100$, $|K| \leq 100 \gamma n \Delta$.
Thus, after removing these $|E(R,\bar{R}) \cup K| \leq \eps n \Delta / 3$ edges, the graph is still $\eps / 3$-far from being $\mathcal{F}$-minor free.

Assume that $|\calE| \leq \alpha n \Delta$.
We will show that with probability at least $96/100$, we can separate $G$ into clusters by removing at most $\alpha n \Delta \cdot 500 f = \eps n \Delta / 2$ edges. 
Therefore, the resulting graph is $\eps / 2$-far from being $\mathcal{F}$-minor free, and with probability at least $2/3$, the algorithm rejects in \cref{step.algmain}.

\paragraph*{Separating $G$ into clusters.}
\begin{enumerate}
\item As argued above, $|E(R,\bar{R}) \cup K| \leq 200 \gamma n \Delta$ with probability at least $98/100$. Therefore, we can separate $G$ into $G[\bar{R}]$ and $G[R_1], \ldots, G[R_j]$, where $R_1, \ldots, R_j$ is the partition of $R$ into remote clusters. 

\item Next, we remove all the edges in $\calE$ (at most $\alpha n \Delta$).
In order to separate each Voronoi cell into its core clusters we simply remove all the edges between different clusters in the same Voronoi cell.
The number of edges which are incident to singleton clusters are at most $\Delta s$.
Since we removed the edges in $\calE$, for a cluster $A$ which is not a singleton, we have that $E(A, \Vor(A)\setminus A) \leq f$.
Therefore by removing at most $s(\Delta + f)$ edges we separate all the Voronoi cells into clusters.

\item By~\cref{lem:marked}, with probability at least $1-o(1)$, we can separate all the clusters, $A$, for which $c(\partial A)$ does not contain a marked center by removing at most $3s f p^{-1}\ln n$ edges. 

\item The expected number of marked clusters is $s p$, therefore with probability at least $99/100$ the number of marked clusters is at most $100sp$.
Thus, with probability at least $99/100$ the number of pairs $A, B \in \bar{R}$ such that $A$ is a cluster and $B$ is a super cluster is at most $s \cdot 100sp$.
Since we removed all edges in $\mathcal{F}$, we have that $E(A,B) < f$ for each such pair.
Therefore, the number of edges between clusters and super clusters is at most $100fs^2 p$.
\end{enumerate}
Recalling from \cref{lem:bound_s} that $s = \Theta(\alpha n^{2/3}/\ln n)$, we can choose the constants in $\alpha$ and $\gamma$ to be small enough such that 
\begin{equation*}
    200 \gamma n \Delta + \Big[ \alpha n \Delta + \Theta\left(\frac{\alpha n^{2/3}}{\ln n}\right)(\Delta + f) \Big] + \Theta(\alpha n f) + \Theta\left(\frac{\alpha^2 n f}{\ln^2 n}\right)
	\leq \eps n \Delta / 2.
\end{equation*}
Hence, with probability at least $96/100$, we can separate $G$ into clusters and remote clusters by removing at most $\eps n \Delta /2$ edges.
\EPF

\bibliographystyle{plain}
\bibliography{refs}

\begin{thebibliography}{10}

\bibitem{BSS08}
I.~Benjamini, O.~Schramm, and A.~Shapira.
\newblock Every minor-closed property of sparse graphs is testable.
\newblock In {\em Proceedings of the Fortieth Annual ACM Symposium on Theory of
  Computing (STOC)}, pages 393--402, 2008.

\bibitem{BodLin93}
H.~L. Bodlaender.
\newblock On {{Linear Time Minor Tests}} with {{Depth}}-{{First Search}}.
\newblock {\em Journal of Algorithms}, 14(1):1--23, 1993.

\bibitem{Bod96}
H.~L. Bodlaender.
\newblock A linear-time algorithm for finding tree-decompositions of small
  treewidth.
\newblock {\em SIAM Journal on Computing}, 25(6):1305--1317, 1996.

\bibitem{Courcelle90}
B.~Courcelle.
\newblock The monadic second-order logic of graphs. i. recognizable sets of
  finite graphs.
\newblock {\em Information and Computation}, 85(1):12--75, 1990.

\bibitem{CGRSSS14}
A.~Czumaj, O.~Goldreich, D.~Ron, C.~Seshadhri, A.~Shapira, and C.~Sohler.
\newblock Finding cycles and trees in sublinear time.
\newblock {\em Random Structure and Algorithms}, 45(2):139--184, 2014.

\bibitem{CzuTes09}
A.~Czumaj, A.~Shapira, and C.~Sohler.
\newblock Testing {{Hereditary Properties}} of {{Nonexpanding
  Bounded}}-{{Degree Graphs}}.
\newblock {\em SIAM Journal on Computing}, 38(6):2499--2510, 2009.

\bibitem{EN17}
M.~Elkin and O.~Neiman.
\newblock Efficient algorithms for constructing very sparse spanners and
  emulators.
\newblock In {\em Proceedings of the Twenty-Eighth Annual {ACM-SIAM} Symposium
  on Discrete Algorithms (SODA)}, pages 652--669, 2017.

\bibitem{ErdCom09}
P.~{Erd\H{o}s} and G.~{Szekeres}.
\newblock {A combinatorial problem in geometry.}
\newblock {\em {Compositio Mathematica}}, 2:463--470, 1935.

\bibitem{GolInt10}
O.~Goldreich.
\newblock Introduction to testing graph properties.
\newblock In {\em Property Testing}, pages 105--141. {Springer}, 2010.

\bibitem{GolInt17a}
O.~Goldreich.
\newblock {\em Introduction to Property Testing}.
\newblock Cambridge University Press, 2017.

\bibitem{GGR98}
O.~Goldreich, S.~Goldwasser, and D.~Ron.
\newblock Property testing and its connection to learning and approximation.
\newblock {\em Journal of the ACM}, 45(4):653--750, 1998.

\bibitem{GolSub99}
O.~Goldreich and D.~Ron.
\newblock {A sublinear bipartiteness tester for bounded degree graphs.}
\newblock {\em {Combinatorica}}, 19(3):335--373, 1999.

\bibitem{GR02}
O.~Goldreich and D.~Ron.
\newblock Property testing in bounded degree graphs.
\newblock {\em Algorithmica}, 32(2):302--343, 2002.

\bibitem{HKNO09}
A.~Hassidim, J.~A. Kelner, H.~N. Nguyen, and K.~Onak.
\newblock Local graph partitions for approximation and testing.
\newblock In {\em Proceedings of the Fiftieth Annual Symposium on Foundations
  of Computer Science (FOCS)}, pages 22--31, 2009.

\bibitem{2018arXiv180508187K}
A.~Kumar, C.~Seshadhri, and A.~Stolman.
\newblock Finding forbidden minors in sublinear time: a o(n\({}^{\mbox{1/2 +
  o(1)}}\))-query one-sided tester for minor closed properties on bounded
  degree graphs.
\newblock {\em arXiv:1805.08187}, 2018.

\bibitem{Kuratowski1930}
C.~Kuratowski.
\newblock Sur le problème des courbes gauches en topologie.
\newblock {\em Fundamenta Mathematicae}, 15(1):271--283, 1930.

\bibitem{LL17}
C.~Lenzen and R.~Levi.
\newblock A local algorithm for the sparse spanning graph problem.
\newblock {\em arXiv:1703.05418}, 2017.

\bibitem{LR15}
R.~Levi and D.~Ron.
\newblock A quasi-polynomial time partition oracle for graphs with an excluded
  minor.
\newblock {\em {ACM} Trans. Algorithms}, 11(3):24:1--24:13, 2015.

\bibitem{RobertsonST94}
N.~Robertson, P.~Seymour, and R.~Thomas.
\newblock Quickly excluding a planar graph.
\newblock {\em Journal of Combinatorial Theory, Series B}, 62(2):323--348,
  1994.

\bibitem{RS04}
N.~Robertson and P.~D. Seymour.
\newblock Graph minors. {XX}. {W}agner's conjecture.
\newblock {\em Journal of Combinatorial Theory, Series B}, 92(2):325--357,
  2004.

\bibitem{Wag37}
K.~Wagner.
\newblock {\"U}ber eine {E}igenschaft der ebenen {K}omplexe.
\newblock {\em Mathematische Annalen}, 114(1):570--590, 1937.

\bibitem{YI15}
Y.~Yoshida and H.~Ito.
\newblock Testing outerplanarity of bounded degree graphs.
\newblock {\em Algorithmica}, 73(1):1--20, 2015.

\end{thebibliography}

\end{document}